\documentclass[twocolumn,english]{IEEEtran}
\usepackage[T1]{fontenc}
\usepackage{color}
\usepackage{babel}
\usepackage{amsmath}
\usepackage{amssymb}
\usepackage{stackrel}
\usepackage{graphicx}
\usepackage[unicode=true,
 bookmarks=true,bookmarksnumbered=true,bookmarksopen=true,bookmarksopenlevel=1,
 breaklinks=false,pdfborder={0 0 0},pdfborderstyle={},backref=false,colorlinks=false]
 {hyperref}
\hypersetup{pdftitle={Your Title},
 pdfauthor={Your Name},
 pdfpagelayout=OneColumn, pdfnewwindow=true, pdfstartview=XYZ, plainpages=false}
\usepackage{breakurl}

\makeatletter
\ifCLASSOPTIONcompsoc
\usepackage[caption=false,font=normalsize,labelfont=sf,textfont=sf]{subfig}
\else
\usepackage[caption=false,font=footnotesize]{subfig}
\fi
\usepackage{cite}
\usepackage{amsthm}
\newtheorem{theorem}{\textbf{Theorem}}
\newtheorem{corollary}{\textbf{Corollary}}
\newtheorem{lemma}{\textbf{Lemma}}
\newtheorem{proposition}{\textbf{Proposition}}
\usepackage{algorithm}
\usepackage{amsmath}

\@ifundefined{showcaptionsetup}{}{%
 \PassOptionsToPackage{caption=false}{subfig}}
\usepackage{subfig}
\makeatother

\begin{document}

\title{\textcolor{black}{The Impact of Antenna Height Difference on the
Performance of Downlink Cellular Networks}}

\author{\IEEEauthorblockN{Junyu Liu$^{1}$, Min Sheng$^{1}$, Kan Wang$^{2}$,
Jiandong Li$^{1}$}\\
\IEEEauthorblockA{$^{1}$State Key Laboratory of Integrated Service
Networks, Xidian University, Xi'an, Shaanxi, 710071, China\\
$^{2}$School of Computer Science and Technology, Xi'an University
of Technology, Xi'an, Shaanxi, 710048, China\\
Email: junyuliu@xidian.edu.cn, \{msheng, jdli\}@mail.xidian.edu.cn,
kanwangkw@outlook.com}}
\maketitle
\begin{abstract}
Capable of significantly reducing cell size and enhancing spatial
reuse, network densification is shown to be one of the most dominant
approaches to expand network capacity. Due to the scarcity of available
spectrum resources, nevertheless, the over-deployment of network infrastructures,
e.g., cellular base stations (BSs), would strengthen the inter-cell
interference as well, thus in turn deteriorating the system performance.
On this account, we investigate the performance of downlink cellular
networks in terms of user coverage probability (CP) and network spatial
throughput (ST), aiming to shed light on the limitation of network
densification. Notably, it is shown that both CP and ST would be degraded
and even diminish to be zero when BS density is sufficiently large,
provided that practical antenna height difference (AHD) between BSs
and users is involved to characterize pathloss. Moreover, the results
also reveal that the increase of network ST is at the expense of the
degradation of CP. Therefore, to balance the tradeoff between user
and network performance, we further study the critical density, under
which ST could be maximized under the CP constraint. Through a special
case study, it follows that the critical density is inversely proportional
to the square of AHD. The results in this work could provide helpful
guideline towards the application of network densification in the
next-generation wireless networks.
\end{abstract}

\section{Introduction\label{sec:Introduction}}

Among the possible approaches to fulfill the unprecedented capacity
goals of the future wireless networks, network densification has been
shown to be the one with the greatest potential \cite{Network_densification_Ref1}.
The basic principle behind network densification is to deploy base
stations (BSs) or access points (APs) with smaller coverage to enable
local spectrum reuse \cite{Ref_F_Yu_small_cell,Ref_F_Yu_SC}. As such,
mobile users are served with short-distance transmission links, thereby
facilitating enormous spatial multiplexing gain and enhancing network
capacity. The benefits of network densification are substantially
verified via the experimental results from Qualcomm \cite{UDN_benefit_Qualcomm}.
Specifically, it is shown that over 1000-fold network capacity gain
can be harvested by deploying 144 self-organizing small cells into
one macro-cell, as compared to the macro-only case. Despite the merits,
however, the experimental results in \cite{UDN_benefit_Qualcomm}
also show that the benefits of network densification in terms of network
capacity enhancement start to diminish when the number of deployed
small cells is sufficiently large. In other words, network densification
may gradually drain the spatial multiplexing gain as well. Therefore,
the limitation of network densification remains to be fully explored.

The research on how network densification impacts the capacity of
wireless networks has received extensive attention in the literature.
In \cite{UPM_Ref_Original,UPM_Ref_K_tier}, the performance of single-tier
cellular networks and multi-tier heterogeneous networks has been investigated,
respectively. Remarkably, it is shown that the network spatial throughput
(ST), an important indicator of network capacity, would linearly increase
with the densification of cellular BSs in both single- and multi-tier
networks. As an encouraging result, it indicates that the potential
spatial multiplexing gain can be sustainably achieved provided that
a sufficient number of BSs are deployed. Nevertheless, the analysis
in \cite{UPM_Ref_Original,UPM_Ref_K_tier} is made on the premise
that only non-line-of-sight (NLOS) paths exist between the transmitters
(Tx's) and the intended receivers (Rx's). Due to the shorter transmission
distance in dense deployment, line-of-sight (LOS) paths are more likely
to appear as well. On this account, authors in \cite{MUPM_Ref_LOS_NLOS,MUPM_Ref_Original,MUPM_Ref_LOS_Conf}
have captured the impact of LOS/NLOS transmissions on the performance
of downlink cellular networks. In particular, it has been observed
that the user coverage probability (CP) tends to decay at some density
and network ST grows sublinearly or even decreases with the increase
of BS density\cite{MUPM_Ref_LOS_Conf}. This is mainly due to the
fact that the inter-cell interference power is likely to overwhelm
the desired signal power when LOS paths exist between interfering
BSs and the intended downlink user. Especially, when BS density further
increases, more interfering BSs would have LOS paths to the intended
user, thereby degrading user and system performance. The results reveal
the limitation of network densification. Furthermore, besides the
scaling law analysis, authors in \cite{MUPM_Ref_LOS_Journal} have
quantified the density, beyond which network ST experiences a notable
decrease.

In the aforementioned research, the 2-D distance is applied to approximate
the distance between the antennas of Tx's and Rx's. In sparsely deployed
networks where Tx's and Rx's are far from each other, such approximation
is of high accuracy and thus valid. When Tx's and Rx's are in proximity,
however, it is apparent that the approximation will lose the accuracy
(see Fig. \ref{fig:scenario}). Hence, it is of great importance to
investigate the performance of ultra-dense networks (UDN) with antenna
height difference (AHD) of Tx's and Rx's. Besides, it is shown from
\cite{MUPM_Ref_LOS_NLOS,MUPM_Ref_Original,MUPM_Ref_LOS_Conf,MUPM_Ref_LOS_Journal}
that the increase of network capacity (system performance) is at the
cost of the deterioration of user performance (e.g., CP). Since user
performance is an important indicator to evaluate the performance
of network densification, it is crucial to balance the tradeoff between
user and network performance.

Motivated by above discussions, we investigate the fundamentals of
network densification in downlink cellular networks with the aid of
stochastic geometry. To explore the impact of AHD between BSs and
downlink users, we study the scaling laws of both CP (user performance)
and network ST (system performance) under a generalized multi-slope
pathloss model (MSPM). Surprisingly, it is shown that, considering
AHD, both CP and network ST would be degraded by network over-densification
and even asymptotically approach zero when BS density is sufficiently
large. The results are opposite to that derived without considering
AHD \cite{MUPM_Ref_LOS_NLOS,MUPM_Ref_Original,MUPM_Ref_LOS_Conf,MUPM_Ref_LOS_Journal}.
Moreover, to guarantee the quality of service (QoS) of users, we further
analyze the critical density that could maximize the network ST under
the CP constraint. It is observed that the critical density is much
smaller (e.g., 10\% or even less under the typical settings) than
the density, under which network ST is maximized without the CP constraint.
The above results could provide helpful insights and guidelines towards
the planning and deployment of future wireless networks.

For the remainder of this paper, we first describe the system model
in Section \ref{sec:System-Model}, followed by a preliminary analysis
on CP and ST under a multi-slope pathloss model in Section \ref{sec:Preliminary Analysis}.
Afterward, we study the CP and ST scaling laws in Section \ref{sec:Scaling Law}
and investigate the critical density under the CP constraint in Section
\ref{sec:Critical Density}. Finally, conclusions are given in Section
\ref{sec:Conclusion}.

\section{System Model\label{sec:System-Model}}

\subsection{Network Model}

\begin{figure}[t]
\begin{centering}
\includegraphics[width=3in]{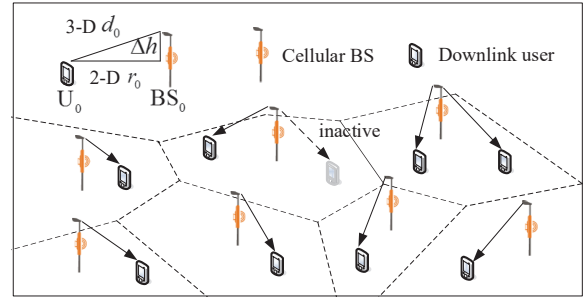}
\par\end{centering}
\caption{\label{fig:scenario}Illustration of downlink cellular networks. Downlink
users are connected to the geometrically nearest BSs. When BSs are
associated with more than one user, one of them are randomly selected
by BSs to serve. Instead of the 2-D distance $r_{i}$ between BSs
and downlink users, the 3-D distance $d_{i}$ between the antennas
of them is considered, involving the AHD $\Delta h$. As an example,
$d_{0}=\sqrt{r_{0}^{2}+\Delta h^{2}}$ for typical downlink $\mathrm{BS}_{0}$-$\mathrm{U}_{0}$.}
\end{figure}

Consider a\textcolor{black}{{} downlink cell network (see Fig. \ref{fig:scenario}),
where BSs (with constant transmit power $P$) and downlink users are
distributed in a two-dimension plane $\mathbb{R}^{2}$, in line with
two independent Homogeneous Poisson Point Processes (HPPPs), $\Pi_{\mathrm{BS}}=\left\{ \mathrm{BS}_{i}\left|\mathrm{BS}_{i}\in\mathbb{R}^{2}\right.\right\} $
and $\Pi_{\mathrm{U}}=\left\{ \mathrm{U}_{j}\left|\mathrm{U}_{j}\in\mathbb{R}^{2}\right.\right\} $
$\left(i,\:j\in\mathbb{N}\right)$, respectively. It is assumed that
all the BSs (downlink users) are equipped with antennas of identical
heights. Meanwhile, denote $\Delta h$ as the AHD between BSs and
users. Downlink users are associated with the geometrically nearest
BSs so as to obtain the strongest average signal strength. It is assumed
that the user density $\lambda_{\mathrm{U}}$ is much greater than
the BS density $\lambda$, i.e., $\lambda_{\mathrm{U}}\gg\lambda$,
to ensure that all the BSs are connected and activated. In each time
slot, BSs would randomly select one of the associated users to serve.
Besides, a saturated data model is considered such that users always
require data to download from the serving BSs.}

Channel power gain consists of two components: pathloss and small-scale
fading. To comprehensively characterize the impact of LOS and NLOS
components, we have adopted an MSPM, i.e.,
\begin{equation}
l_{N}\left(\left\{ \alpha_{n}\right\} _{n=0}^{N-1};x\right)=K_{n}x^{-\alpha_{n}},\:R_{n}\leq x<R_{n+1}\label{eq:MUPM}
\end{equation}
where $K_{0}=1$, $K_{n}=\prod_{i=1}^{n}R_{i}^{\alpha_{i}-\alpha_{i-1}}$
$\left(n\geq1\right)$, $0=R_{0}<R_{1}<\cdots<R_{N}=\infty$ and $0\leq\alpha_{0}\leq\alpha_{1}\leq\cdots\leq\alpha_{N-1}$
($\alpha_{N-1}>2$ for practical concerns \cite{MUPM_Ref_Original})
.

From (\ref{eq:MUPM}), it follows that different pathloss exponents
are used to characterize the attenuation rates of signal power within
different regions. For instance, when $N=2$, MSPM degenerates into
the dual-slope pathloss model (DSPM) \cite{MUPM_Ref_Original,MUPM_Ref_K_dimension}
\begin{equation}
l_{2}\left(\alpha_{0},\alpha_{1};x\right)=\left\{ \begin{array}{cc}
x^{-\alpha_{0}}, & x\leq R_{1}\\
K_{1}x^{-\alpha_{1}}, & x>R_{1}
\end{array}\right.\label{eq:DUPM}
\end{equation}
where $K_{1}=R_{1}^{\alpha_{1}-\alpha_{0}}$. The DSPM in (\ref{eq:DUPM})
is applied when an LOS path and a ground-reflected path exist between
Tx and the intended Rx. As such, signal power attenuates slowly (with
rate $\alpha_{0}$) within a \textit{corner distance} $R_{1}$, while
attenuates much more quickly (with rate $\alpha_{1}$) with distance
out of $R_{1}$. When $N=1$, MSPM further degenerates into the most
widely used single-slope pathloss model (SSPM) \cite{UPM_Ref_Original,MUPM_Ref_K_dimension}
\begin{equation}
l_{1}\left(\alpha_{0};x\right)=x^{-\alpha_{0}},\:x\in\left[0,\infty\right).\label{eq:SUPM}
\end{equation}

For small-scale fading, although it is more suitable to use Rice fading
when LOS paths exist between Tx's and Rx's, insightful results could
hardly be obtained due to the complicated form. Instead, Rayleigh
fading with zero mean and unit variance $h\sim\mathcal{CN}\left(0,1\right)$
is applied to model small-scale fading for mathematical tractability.
In additional, as will be shown in Section \ref{sec:Scaling Law},
the application of Rayleigh fading will not impact the results on
CP and ST scaling laws via the comparison between numerical and simulation
results.

\subsection{Performance Metrics}

We adopt CP and ST to reflect user performance and system performance,
respectively. To be specific, following the signal-to-interference
ratio (SIR) at the typical downlink user $\mathrm{U}_{0}$\footnote{Without loss of generality, we evaluate the CP of downlink pair $\mathrm{BS}_{0}$-$\mathrm{U}_{0}$.
Meanwhile, as spectrum resources could be universally reused, inter-cell
interference dominates the performance of downlink networks. Hence,
the impact of noise is ignored. }, CP is defined as
\begin{equation}
\mathsf{CP}\left(\lambda\right)=\mathbb{P}\left\{ \mathsf{SIR}_{\mathrm{U}_{0}}>\tau\right\} ,\label{eq: define CP}
\end{equation}
where $\tau$ denotes the decoding threshold. Based on CP in (\ref{eq: define CP}),
we further define network ST as
\begin{equation}
\mathsf{ST}\left(\lambda\right)=\lambda\mathbb{P}\left\{ \mathsf{SIR}_{\mathrm{U}_{0}}>\tau\right\} \log_{2}\left(1+\tau\right),\:\left[\mathrm{bits}/\left(\mathrm{s\cdot Hz\cdot m^{2}}\right)\right]\label{eq: define ST}
\end{equation}
which could characterize the number of bits that are successfully
conveyed over unit time, frequency and area. Hence, ST serves as an
indicator to network capacity.

\textbf{Notation}: If $_{2}F_{1}\left(\cdot,\cdot,\cdot,\cdot\right)$
is defined as the standard Gaussian hypergeometric function, denote
$\omega_{1}\left(x,\alpha_{n}\right)=$ $_{2}F_{1}\left(1,1-\frac{2}{\alpha_{n}},2-\frac{2}{\alpha_{n}},-x\right)$
and $\omega_{2}\left(x,\alpha_{n}\right)=$ $_{2}F_{1}\left(1,\frac{2}{\alpha_{n}},1+\frac{2}{\alpha_{n}},-x\right)$
in the rest of the paper.

\section{Analysis of CP and ST with AHD\label{sec:Preliminary Analysis}}

In this section, we first give preliminary analysis of CP and ST under
the MSPM in (\ref{eq:MUPM}). Particularly, the impact of the AHD
on the network performance is highlighted.

From (\ref{eq: define CP}), CP is defined based on the SIR evaluated
at $\mathrm{U}_{0}$. Therefore, we first characterize the SIR at
$\mathrm{U}_{0}$ as
\begin{align}
\mathsf{SIR}_{\mathrm{U}_{0}}= & PH_{\mathrm{U}_{0},\mathrm{BS}_{0}}l_{N}\left(\left\{ \alpha_{n}\right\} _{n=0}^{N-1};d_{0}\right)/I_{\mathrm{IC}},\label{eq:define SIR}
\end{align}
where $I_{\mathrm{IC}}=\underset{\tiny{\mathrm{BS}_{i}\in\tilde{\Pi}_{\mathrm{BS}}}}{\sum}PH_{\mathrm{U}_{0},\mathrm{BS}_{i}}l_{N}\left(\left\{ \alpha_{n}\right\} _{n=0}^{N-1};d_{i}\right)$
denotes the inter-cell interference suffered by $\mathrm{U}_{0}$,
$\tilde{\Pi}_{\mathrm{BS}}=\Pi_{\mathrm{BS}}\backslash\mathrm{BS}_{0}$,
$d_{i}$ denotes the distance from the antenna of $\mathrm{BS}_{i}$
to that of $\mathrm{U}_{0}$, and $H_{\mathrm{U}_{0},\mathrm{BS}_{i}}$
denotes the corresponding channel power gain caused by small-scale
fading. Meanwhile, if we denote $r_{i}$ as the distance from $\mathrm{BS}_{i}$
to $\mathrm{U}_{0}$, then $d_{i}=\sqrt{r_{i}^{2}+\Delta h^{2}}$.
Note that $H_{\mathrm{U}_{0},\mathrm{BS}_{i}}\sim\exp\left(1\right)$
since Rayleigh fading $h\sim\mathcal{CN}\left(0,1\right)$ is applied
to model small-scale fading.

\begin{figure*}[t]
\begin{equation}
\mathsf{CP}_{N}\left(\lambda\right)=\left\{ \begin{array}{cc}
\frac{1}{1+C_{1}}\exp\left(-\pi\lambda C_{1}\bigtriangleup h^{2}\right), & N=1\\
\stackrel[n=0]{N-1}{\sum}\mathbb{E}_{r_{0}\in\left[R_{n},R_{n+1}\right)}\left\{ \exp\left[-\pi\lambda\left(\bar{R}_{n+1}^{2}\omega_{2}\left(\frac{\bar{R}_{n+1}^{\alpha_{n}}}{\tau d_{0}^{\alpha_{n}}},\alpha_{n}\right)-d_{0}^{2}\omega_{2}\left(\tau^{-1},\alpha_{n}\right)\right.\right.\right.\\
+\left.\left.\left.\stackrel[i=n+1]{N-1}{\sum}\left(\bar{R}_{i+1}^{2}\omega_{2}\left(\frac{\bar{R}_{i+1}^{\alpha_{i}}}{\tau K_{i}d_{0}^{\alpha_{n}}},\alpha_{i}\right)-\bar{R}_{i}^{2}\omega_{2}\left(\frac{\bar{R}_{i}^{\alpha_{i}}}{\tau K_{i}d_{0}^{\alpha_{n}}},\alpha_{i}\right)\right)\right)\right]\right\} , & N>1
\end{array}\right.\label{eq:CP general}
\end{equation}
\end{figure*}

From (\ref{eq:define SIR}), we can obtain the following results on
CP and ST in Proposition \ref{proposition: CP and ST}.

\begin{proposition}

Considering the AHD between BSs and downlink users, the ST in downlink
cellular networks under MSPM in (\ref{eq:MUPM}) is given by $\mathsf{ST}_{N}\left(\lambda\right)=\lambda\mathsf{CP}_{N}\left(\lambda\right)\log_{2}\left(1+\tau\right)$,
where $\mathsf{CP}_{N}\left(\lambda\right)$ is given by (\ref{eq:CP general})
at the top of Page 4. In (\ref{eq:CP general}), $C_{1}=\frac{2\tau\omega_{1}\left(\tau,\alpha_{0}\right)}{\alpha_{0}-2}$,
$d_{0}=\sqrt{r_{0}^{2}+\Delta h^{2}}$ and the probability density
function (PDF) of $r_{0}$ is derived from the contact distribution
\cite{book_stochastic_geometry}
\begin{equation}
f_{r_{0}}\left(x\right)=2\pi\lambda x\exp\left(-\pi\lambda x^{2}\right),\:x\geq0.\label{eq:PDF of r0}
\end{equation}

\label{proposition: CP and ST}

\end{proposition}

\textit{Proof}: Please refer to Appendix \ref{subsec:Proof for SP and ST original}.\qed

Despite its complicated form, the result in Proposition \ref{proposition: CP and ST}
could provide a numerical approach to capture the relationship between
system parameters and performance metrics, namely, CP and ST, under
MSPM. Meanwhile, according to the special case in (\ref{eq:CP general}),
where $N=1$, it follows that both CP and ST would exponentially decrease
with $\Delta h^{2}$. In other words, the results, without considering
the impact of the AHD, greatly over-estimate the performance of downlink
network. In addition, when $N=2$ and MSPM degenerates into DSPM,
the results on CP and ST could be further simplified according to
the following corollary.

\begin{corollary}

Considering the AHD between BSs and downlink users, the ST in downlink
cellular networks under DSPM in (\ref{eq:DUPM}) is given by $\mathsf{ST}_{2}\left(\lambda\right)=\lambda\mathsf{CP}_{2}\left(\lambda\right)\log_{2}\left(1+\tau\right)$,
where
\begin{align}
\mathsf{CP}_{2}\left(\lambda\right)= & \mathbb{E}_{r_{0}\in\left[0,R_{1}\right)}\left[e^{-\pi\lambda\left(\delta_{1}\left(\alpha_{0},d_{0},\tau,R_{1}\right)+\delta_{2}\left(\alpha_{0},\alpha_{1},d_{0},\tau,R_{1}\right)\right)}\right]\nonumber \\
+ & \mathbb{E}_{r_{0}\in\left[R_{1},\infty\right)}\left[e^{-\pi\lambda\delta_{3}\left(\alpha_{1},d_{0},\tau\right)}\right].\label{eq:CP DUPM}
\end{align}
In (\ref{eq:CP DUPM}), $d_{0}=\sqrt{r_{0}^{2}+\Delta h^{2}}$, $\delta_{1}\left(\alpha_{0},d_{0},\tau,R_{1}\right)=R_{1}^{2}\omega_{2}\left(\frac{R_{1}^{\alpha_{0}}}{\tau d_{0}^{\alpha_{0}}},\alpha_{0}\right)-d_{0}^{2}\omega_{2}\left(\frac{1}{\tau},\alpha_{0}\right)$,
$\delta_{2}\left(\alpha_{0},\alpha_{1},d_{0},\tau,R_{1}\right)=\frac{2\tau d_{0}^{\alpha_{0}}R_{1}^{2-\alpha_{0}}}{\alpha_{1}-2}\omega_{1}\left(\frac{\tau d_{0}^{\alpha_{0}}}{R_{1}^{\alpha_{0}}},\alpha_{1}\right)$,
$\delta_{3}\left(\alpha_{1},d_{0},\tau\right)=\frac{2\tau d_{0}^{2}}{\alpha_{1}-2}\omega_{1}\left(\tau,\alpha_{1}\right)$
and the PDF of $r_{0}$ is given by (\ref{eq:PDF of r0}).

\label{corollary: CP and ST DUPM}

\end{corollary}

\textit{Proof}: The proof can be completed by setting $N=2$ in (\ref{eq:CP general})
with easy manipulation, and thus omitted due to space limitation.\qed

\begin{figure}[t]
\begin{centering}
\subfloat[CP.]{\begin{centering}
\includegraphics[width=3.5in]{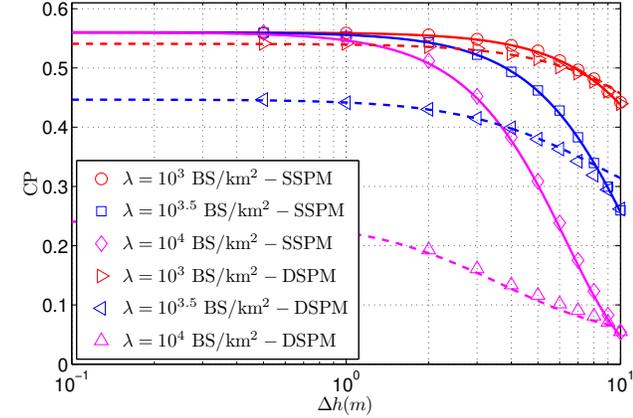}
\par\end{centering}
}
\par\end{centering}
\begin{centering}
\subfloat[ST.]{\begin{centering}
\includegraphics[width=3.5in]{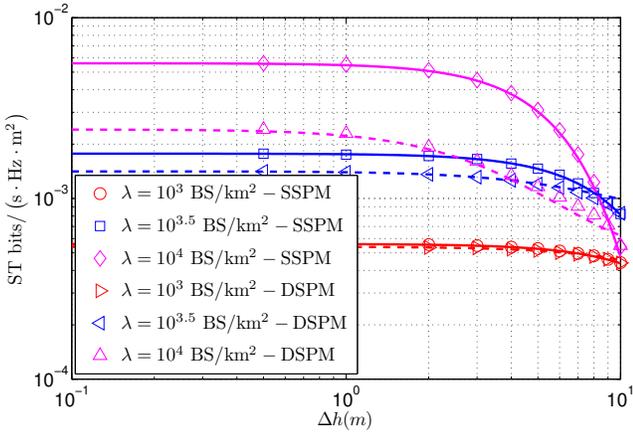}
\par\end{centering}
}
\par\end{centering}
\caption{\label{fig:CP and ST with AHD}CP and ST varying with AHD $\Delta h$.
For system settings, set $P=23$dBm and $\tau=0$dB. For SSPM, set
$\alpha_{0}=4$. For DSPM, set $\alpha_{0}=1.5$, $\alpha_{1}=4$
and $R_{1}=10$m. Lines and markers denote numerical and simulation
results, respectively, in this figure and the remaining figures in
this paper.}
\end{figure}

Based on Proposition \ref{proposition: CP and ST} and Corollary \ref{corollary: CP and ST DUPM},
we illustrate the impact of AHD on CP and network ST in detail. In
particular, Fig. \ref{fig:CP and ST with AHD} shows the CP and ST
as a function of $\Delta h$ of BSs and downlink users under different
BS densities. It can be seen from Fig. \ref{fig:CP and ST with AHD}
that both CP and ST would be degraded by $\Delta h$. This indicates
that, although the existence of $\Delta h$ would weaken both desired
and interference signal power, the decrease of the desired signal
power overwhelms that of the interference signal powers. Meanwhile,
it is shown that the impact of $\Delta h$ on CP and ST is relatively
small under sparse BS deployment, while the impact is significant
under dense BS deployment. Hence, in dense wireless networks, where
the user antenna heights are basically small, it is preferable to
deploy small cell BSs with smaller antenna heights so as to reduce
the AHD, thereby ensuring the user performance as well as system performance.

As shown in Fig. \ref{fig:CP and ST with AHD}, it is evident that
the existence of $\Delta h$ leads to the performance degradation
in terms of CP and ST, especially in the fully densified networks.
Therefore, we have to further explore the influence of $\Delta h$
on the scaling laws of CP and ST in the following.

\section{CP and ST Scaling Laws\label{sec:Scaling Law}}

In this part, before investigating the fundamental limitation of network
densification by analyzing the CP and ST scaling laws, results on
$\omega_{1}\left(x,y\right)$ are first given in the following Lemma.

\begin{lemma}

For $y>2$, $\omega_{1}\left(x,y\right)$ is a decreasing function
of $x$.

\label{lemma: hypergeometric function}

\end{lemma}

\textit{Proof}: Please refer to the proof for Lemma 1 in \cite{Ref_SBPM}.\qed

On the basis of Lemma \ref{lemma: hypergeometric function} and Proposition
\ref{proposition: CP and ST}, we show the CP and ST scaling laws
in Theorem \ref{theorem: CP and ST scaling law}.

\begin{theorem}

When AHD exists between BSs and downlink users, CP and ST scale with
BS density $\lambda$ as $\mathsf{CP}_{N}\left(\lambda\right)\sim e^{-\kappa\lambda}$
and $\mathsf{ST}_{N}\left(\lambda\right)\sim\lambda e^{-\kappa\lambda}$
($\kappa$ is a constant), respectively, under MSPM.

\label{theorem: CP and ST scaling law}

\end{theorem}

\textit{Proof}: Please refer to Appendix \ref{subsec:Proof for scaling law}.\qed

\begin{figure}[t]
\begin{centering}
\subfloat[\label{fig:CP scaling law}CP.]{\begin{centering}
\includegraphics[width=3.2in]{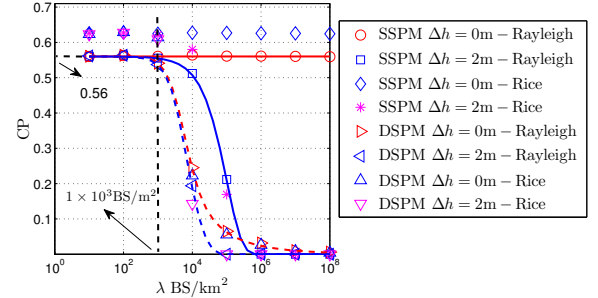}
\par\end{centering}
}
\par\end{centering}
\begin{centering}
\subfloat[\label{fig:ST scaling law}ST.]{\begin{centering}
\includegraphics[width=3.2in]{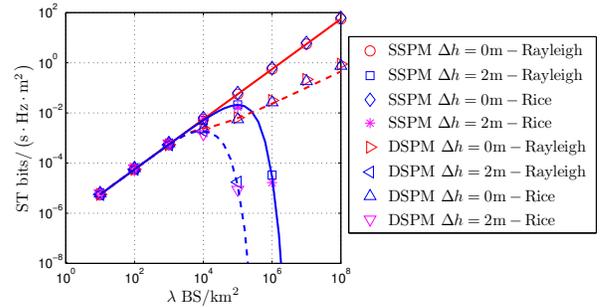}
\par\end{centering}
}
\par\end{centering}
\caption{\label{fig:CP and ST scaling law}CP and ST varying with BS density
$\lambda$. For system settings, set $P=23$dBm and $\tau=0$dB. For
SSPM, set $\alpha_{0}=4$. For DSPM, set $\alpha_{0}=1.5$, $\alpha_{1}=4$
and $R_{1}=10$m. To reflect the impact of LOS paths on signal propagation,
we set $\upsilon_{\mathrm{NC}}=1$ and $\upsilon_{\mathrm{DoF}}=12$
for Rice fading.}
\end{figure}

It is shown from Theorem \ref{theorem: CP and ST scaling law} that
both user and system performance would be degraded when BS density
is sufficiently large. This is essentially different from the results
in \cite{MUPM_Ref_LOS_NLOS,MUPM_Ref_Original,MUPM_Ref_LOS_Conf,MUPM_Ref_LOS_Journal},
where the impact of AHD has not been taken into account in the scaling
law analysis. Particularly, we show the difference in Fig. \ref{fig:CP and ST scaling law}.

Fig. \ref{fig:CP and ST scaling law} shows the CP and ST as a function
of BS density $\lambda$ under different $\Delta h$. It is shown
in Fig. \ref{fig:CP scaling law} that, when $\Delta h=0$m, CP almost
keeps constant with the increasing $\lambda$ under SSPM, and slowly
decreases with the increasing $\lambda$ under DSPM (compared to the
DSPM case under $\Delta h>0$m). In consequence, network ST would
linearly/sublinearly grow with $\lambda$, as shown in Fig. \ref{fig:ST scaling law}.
In contrast, both CP and ST asymptotically approach zero when $\lambda$
is sufficiently large under $\Delta h>0$m. In practice, AHD would
exist between BSs and cellular users, even when small cell BSs are
densely deployed. Therefore, the results, which ignore the impact
of AHD, have over-estimated the benefits of network densification,
while those in Theorem \ref{theorem: CP and ST scaling law} could
shed light on the fundamental limitation of network densification.

To verify the validity of the scaling law analysis under Rayleigh
fading, we also evaluate the performance of downlink networks under
Rice fading via simulation results in Fig. \ref{fig:CP and ST scaling law}.
Specifically, the channel power gain under Rice fading channels follows
the non-central $\chi^{2}$ distribution with non-centrality parameter
$\upsilon_{\mathrm{NC}}$ and degrees of freedom $\upsilon_{\mathrm{DoF}}$.
A larger $\upsilon_{\mathrm{DoF}}$ indicates more scattered components.
It can be seen from Fig. \ref{fig:CP and ST scaling law} that, although
gaps exist between the results under Rice and Rayleigh fadings, it
is apparent that the CP and ST scaling laws under Rice fading are
identical as those under Rayleigh fading.

In addition, it is observed from Fig. \ref{fig:CP and ST scaling law}
that the improvement of system performance is at the cost of the degeneration
of user experience. For instance, when $\Delta h>0$m, network ST
grows with BS density at $\lambda=1\times10^{3}\mathrm{BS}/\mathrm{km}^{2}$
(see Fig. \ref{fig:CP scaling law}), under which CP already starts
to diminish with $\lambda$ (see Fig. \ref{fig:ST scaling law}).
Therefore, besides ensuring the system performance, it is also critical
to guarantee the user experience when planning the deployment of cellular
networks, the detail of which will be described in the next section.

\section{Critical Density Under CP Constraint\label{sec:Critical Density}}

In this section, a CP requirement $\varepsilon$ is set to guarantee
the QoS of users as
\begin{equation}
\mathsf{CP}\left(\lambda\right)=\mathbb{P}\left\{ \mathsf{SIR}_{\mathrm{U}_{0}}>\tau\right\} >\varepsilon.\label{eq: define CP with QoS Req}
\end{equation}
From (\ref{eq: define CP with QoS Req}), it is intuitive that whether
or not the constraint could be satisfied greatly depends on the deployment
density of BSs. Nonetheless, as observed from Fig. \ref{fig:CP scaling law},
the maximal CP that can be achieved reaches 0.56, irrespective of
the BS density. Therefore, besides BS density, other parameters such
as pathloss exponents, decoding threshold, etc., may impact whether
the CP requirement can be met as well. In this light, we first analyze
the necessary condition to acquire the CP requirement. Afterward,
we derive the critical density, under which network ST can be maximized
under the pre-set CP requirement. 

It is worth noting that, to provide helpful insights towards the impact
of system parameters on necessary regions and critical density, the
results derived in this section are built on the SSPM in (\ref{eq:SUPM}).
In the following theorem, the results on the necessary condition are
first given.

\begin{theorem}

Under SSPM in (\ref{eq:SUPM}), the necessary condition to satisfy
the CP requirement in (\ref{eq: define CP with QoS Req}) is given
by 
\begin{equation}
\frac{2\tau\omega_{1}\left(\tau,\alpha_{0}\right)}{\alpha_{0}-2}<\varepsilon^{-1}-1.\label{eq: necessary condition}
\end{equation}

\label{theorem: feasible region}

\end{theorem}

\textit{Proof}: Please refer to Appendix \ref{subsec:Proof for Theorem condition}.\qed

Theorem \ref{theorem: feasible region} provides a direct approach
on how to reasonably adjust system parameters to meet the pre-set
CP requirement of downlink users. Meanwhile, the right-hand-side of
(\ref{eq: necessary condition}), i.e., $g\left(\varepsilon\right)=\varepsilon^{-1}-1$,
implies that $g\left(\varepsilon\right)$ exponentially decreases
with $\varepsilon$ and approaches 0 when $\varepsilon\rightarrow1$.
Therefore, it is more difficult to meet the CP requirement especially
when $\varepsilon$ grows larger. Aided by Theorem \ref{theorem: feasible region},
we further obtain the critical BS density in the following corollary.

\begin{corollary}

With the CP constraint $\varepsilon$, the critical BS density $\lambda^{*}$,
under which network ST is maximized, is given by
\begin{equation}
\lambda^{*}=\frac{\left(\alpha_{0}-2\right)\ln\left[\varepsilon^{-1}\left(1+\frac{2\tau\omega_{1}\left(\tau,\alpha_{0}\right)}{\alpha_{0}-2}\right)^{-1}\right]}{2\pi\tau\omega_{1}\left(\tau,\alpha_{0}\right)\bigtriangleup h^{2}}.\label{eq:critical density with Req}
\end{equation}
Without the CP constraint, the critical BS density $\lambda^{\dagger}$,
under which network ST is maximized, is given by
\begin{equation}
\lambda^{\dagger}=\frac{\alpha_{0}-2}{2\pi\tau\omega_{1}\left(\tau,\alpha_{0}\right)\bigtriangleup h^{2}}.\label{eq:critical density wo Req}
\end{equation}

\label{corollary: critical density}

\end{corollary}

\textit{Proof}: It is straightforward to obtain $\lambda^{*}$ following
(\ref{eq:proof for theorem condition}) in Appendix \ref{subsec:Proof for Theorem condition}
and $\lambda^{\dagger}$ by solving $\frac{\partial\mathsf{ST}_{1}\left(\lambda\right)}{\partial\lambda}=0$,
where $\mathsf{ST}_{1}\left(\lambda\right)$ is given by Proposition
\ref{proposition: CP and ST}.\qed

\begin{figure}[t]
\begin{centering}
\includegraphics[width=3.5in]{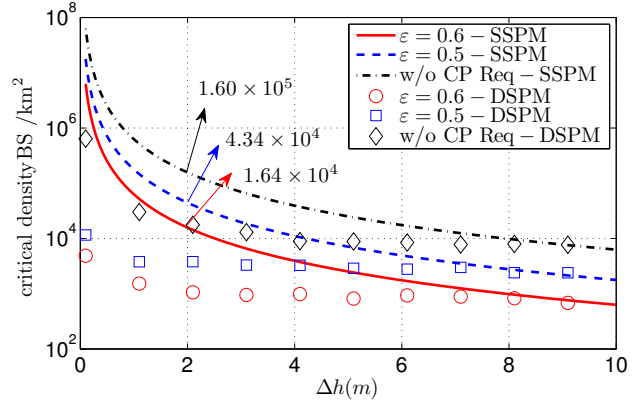}
\par\end{centering}
\caption{\label{fig:critical density}Critical densities $\lambda^{*}$ and
$\lambda^{\dagger}$ varying with the AHD $\Delta h$. For system
settings, set $P=23$dBm and $\tau=0$dB. For SSPM, set $\alpha_{0}=5$.
For DSPM, set $\alpha_{0}=1.5$, $\alpha_{1}=5$ and $R_{1}=10$m.}
\end{figure}

The influence of system parameters on critical densities is captured
using the closed-form expression in Corollary \ref{corollary: critical density}.
Especially, it is observed that both $\lambda^{*}$ and $\lambda^{\dagger}$
are inversely proportional to the square of AHD, i.e., $\Delta h^{2}$.
Meanwhile, we extend the results into the case with DSPM applied.
Specifically, we plot the critical density as a function of $\Delta h$
under both SSPM and DSPM in Fig. \ref{fig:critical density}. Due
to space limitation, numerical results on the critical densities under
DSPM are not presented and only simulation results (drawn by markers)
are given.

We observe from Fig. \ref{fig:critical density} that the CP constraint
greatly limits the maximal deployment density of BSs in downlink networks.
For instance, the critical density is reduced by 3.6 and even 9.7
folds when $\varepsilon=0.5$ and $\varepsilon=0.6$, respectively,
given $\Delta h=2$m under the settings in Fig. \ref{fig:critical density}.
Meanwhile, critical densities $\lambda^{*}$ and $\lambda^{\dagger}$
would exponentially decrease with $\Delta h$ under single-slope and
dual-slope models. Therefore, the above results also reveal the essential
impact of AHD on the BS deployment in downlink cellular network. In
particular, it indicates that, in densely deployed scenarios (e.g,
stadium and open gathering), the antenna height of small cell BSs
should be lowered, thereby facilitating the maximization of network
ST while ensuring the QoS of downlink users.

\section{Conclusion\label{sec:Conclusion}}

In this paper, we have explored the fundamental limits of network
densification in downlink cellular networks under a generalized multi-slope
pathloss model. Specifically, considering the AHD between BSs and
downlink users, it is shown that the network ST first increases, then
decreases with network densification and finally approaches zero when
BSs are over-deployed.  Meanwhile, it is observed that the CP of
downlink users starts to diminish with the BS density when network
ST is increased. Therefore, to strike a balance between user and system
performance, we have derived the critical density, under which network
ST can be maximized with the pre-set CP constraint. The results in
this work could provide helpful guidance for the network deployment
and application of network densification in future wireless networks.

\appendix

\section{*}

\subsection{Proof for Proposition \ref{proposition: CP and ST}\label{subsec:Proof for SP and ST original}}

Substitute (\ref{eq:define SIR}) into (\ref{eq: define CP}), we
have
\begin{align}
\mathsf{CP}_{N}\left(\lambda\right)= & \mathbb{P}\left\{ H_{\mathrm{U}_{0},\mathrm{BS}_{0}}>s_{N}I_{\mathrm{IC}}\right\} \nonumber \\
\overset{\left(\mathrm{a}\right)}{=} & \mathbb{E}_{d_{0},\tilde{\Pi}_{\mathrm{BS}},H_{\mathrm{U}_{0},\mathrm{BS}_{i}}}\left[\underset{\tiny{\mathrm{BS}_{i}\in\tilde{\Pi}_{\mathrm{BS}}}}{\prod}e^{-s_{N}PH_{\mathrm{U}_{0},\mathrm{BS}_{i}}l_{N}\left(d_{i}\right)}\right]\nonumber \\
\overset{\left(\mathrm{b}\right)}{=} & \mathbb{E}_{d_{0},\tilde{\Pi}_{\mathrm{BS}}}\left[\underset{\tiny{\mathrm{BS}_{i}\in\tilde{\Pi}_{\mathrm{BS}}}}{\prod}\frac{1}{1+s_{N}Pl_{N}\left(d_{i}\right)}\right],\label{eq:CP general proof 0}
\end{align}
where $s_{N}=\frac{\tau}{Pl_{N}\left(d_{0}\right)}$. In (\ref{eq:CP general proof 0}),
(a) and (b) are due to $H_{\mathrm{U}_{0},\mathrm{BS}_{i}}\sim\exp\left(1\right)$
and the independence of $H_{\mathrm{U}_{0},\mathrm{BS}_{i}}$. Aided
by the probability generating functional (PGFL) of Poisson point process
(PPP) \cite{book_stochastic_geometry}, $\mathsf{CP}_{N}\left(\lambda\right)$
in (\ref{eq:CP general proof 0}) further turns into
\begin{align}
\mathsf{CP}_{N}\left(\lambda\right)= & \mathbb{E}_{d_{0}}\left[e^{-\lambda\int_{d_{0}}^{\infty}\left(1-\frac{1}{1+s_{N}Pl_{N}\left(x\right)}\right)d\left(\pi x^{2}\right)}\right],\nonumber \\
= & \mathbb{E}_{d_{0}}\left[e^{-2\pi\lambda\int_{d_{0}}^{\infty}x\left(1-\frac{1}{1+s_{N}Pl_{N}\left(x\right)}\right)dx}\right].\label{eq:CP general proof 1}
\end{align}

Given $N=1$, it is straightforward to obtain $s_{1}=\frac{\tau d_{0}^{\alpha_{0}}}{P}$
and
\begin{align}
\mathsf{CP}_{1}\left(\lambda\right)= & \mathbb{E}_{d_{0}}\left[\exp\left(-\frac{2\pi\lambda\tau\omega_{1}\left(\tau,\alpha_{0}\right)}{\alpha-2}d_{0}^{2}\right)\right]\nonumber \\
= & \mathbb{E}_{r_{0}}\left[\exp\left(-\frac{2\pi\lambda\tau\omega_{1}\left(\tau,\alpha_{0}\right)}{\alpha-2}\left(r_{0}^{2}+\Delta h^{2}\right)\right)\right]\nonumber \\
\overset{\left(\mathrm{a}\right)}{=} & \frac{1}{1+C_{1}}\exp\left(-\pi\lambda C_{1}\bigtriangleup h^{2}\right),\label{eq:CP general proof 2}
\end{align}
where (a) follows because the PDF of $r_{0}$ is given by (\ref{eq:PDF of r0}).

Given $N>2$ and $d_{0}\in\left[\bar{R}_{n},\bar{R}_{n+1}\right)$
with $\bar{R}_{n}=\sqrt{r_{0}^{2}+R_{n}^{2}}$, $\int_{d_{0}}^{\infty}x^{k-1}\left(1-\frac{1}{1+s_{N}Pl_{N}\left(x\right)}\right)dx$
in (\ref{eq:CP general proof 1}) turns into
\begin{align*}
 & \int_{d_{0}}^{\infty}x\left(1-\frac{1}{1+s_{N}Pl_{N}\left(x\right)}\right)dx\\
= & \int_{d_{0}}^{\bar{R}_{n+1}}x\left(1-\frac{1}{1+\tau d_{0}^{\alpha_{n}}x^{-\alpha_{n}}}\right)dx\\
+ & \stackrel[i=n+1]{N-1}{\sum}\int_{\bar{R}_{i}}^{\bar{R}_{i+1}}x\left(1-\frac{1}{1+\tau K_{i}d_{0}^{\alpha_{n}}x^{-\alpha_{i}}}\right)dx\\
= & \frac{1}{2}\left[\bar{R}_{n+1}^{2}\omega_{2}\left(\frac{\bar{R}_{n+1}^{\alpha_{n}}}{\tau d_{0}^{\alpha_{n}}},\alpha_{n}\right)-d_{0}^{2}\omega_{2}\left(\tau^{-1},\alpha_{n}\right)\right]\\
+ & \stackrel[i=n+1]{N-1}{\sum}\left[\frac{\bar{R}_{i+1}^{2}}{2}\omega_{2}\left(\frac{\bar{R}_{i+1}^{\alpha_{i}}}{\tau K_{i}d_{0}^{\alpha_{n}}},\alpha_{i}\right)-\frac{\bar{R}_{i}^{2}}{2}\omega_{2}\left(\frac{\bar{R}_{i}^{\alpha_{i}}}{\tau K_{i}d_{0}^{\alpha_{n}}},\alpha_{i}\right)\right]
\end{align*}
Hence, the proof is completed.

\subsection{Proof for Theorem \ref{theorem: CP and ST scaling law}\label{subsec:Proof for scaling law}}

From Proposition \ref{proposition: CP and ST}, it follows that the
proof for the scaling laws of CP and ST under the SSPM is straightforward
and thus omitted due to limitation. Then, we focus on the proof for
the case with $N>2$, and some useful notations are first given in
the following.

Denote $g_{1}\left(x\right)$ and $g_{2}\left(x\right)$ as two functions
on the subset of real numbers. We write $g_{1}\left(x\right)=\Omega\left(g_{2}\left(x\right)\right)$
if $\exists m>0$, $x_{0}$, $\forall x>x_{0}$, $m\left|g_{2}\left(x\right)\right|\leq\left|g_{1}\left(x\right)\right|$,
and $g_{1}\left(x\right)=\mathcal{O}\left(g_{2}\left(x\right)\right)$
if $\exists m>0$, $x_{0}$, $\forall x>x_{0}$, $\left|g_{1}\left(x\right)\right|\leq m\left|g_{2}\left(x\right)\right|$.

Given $N>2$, the CP in (\ref{eq:CP general}) can be expressed as

\begin{align}
 & \mathsf{CP}_{N}\left(\lambda\right)\nonumber \\
= & \mathbb{E}_{r_{0}\in\left[R_{0},R_{N-1}\right)}\left[e^{-2\pi\lambda\int_{d_{0}}^{\infty}x\left(1-\frac{1}{1+s_{N}Pl_{N}\left(x\right)}\right)dx}\right]\nonumber \\
+ & \mathbb{E}_{r_{0}\in\left[R_{N-1},R_{N}\right)}\left[e^{-2\pi\lambda\int_{d_{0}}^{\infty}x\left(1-\frac{1}{1+s_{N}Pl_{N}\left(x\right)}\right)dx}\right].\label{eq:bound proof 1}
\end{align}

Then, it can be directly obtained that
\begin{align}
\mathsf{CP}_{N}\left(\lambda\right)> & \mathbb{E}_{r_{0}\in\left[R_{N-1},R_{N}\right)}\left[e^{-2\pi\lambda\int_{d_{0}}^{\infty}x\left(1-\frac{1}{1+s_{N}Pl_{N}\left(x\right)}\right)dx}\right].\label{eq:bound proof 2}
\end{align}
As $d_{0}=\sqrt{r_{0}^{2}+\Delta h^{2}}$, $\bar{R}_{N-1}=\sqrt{R_{N-1}^{2}+\Delta h^{2}}$
and $R_{N}=\infty$, when $d_{0}\in\left[\bar{R}_{N-1},\infty\right)$,
$s_{N}=\frac{\tau}{PK_{N-1}d_{0}^{-\alpha_{N-1}}}$ and $l_{N}\left(x\right)=K_{N-1}x^{-\alpha_{N-1}}$,
the integral in (\ref{eq:bound proof 2}) turns into
\begin{align}
 & \int_{d_{0}}^{\infty}x\left(1-\frac{1}{1+\tau d_{0}^{\alpha_{N-1}}x^{-\alpha_{N-1}}}\right)dx\nonumber \\
= & \tau\omega_{1}\left(\tau,\alpha_{N-1}\right)d_{0}^{2}\nonumber \\
= & \tau\omega_{1}\left(\tau,\alpha_{N-1}\right)\left(r_{0}^{2}+\Delta h^{2}\right).\label{eq:bound proof 3}
\end{align}
Next, we derive the lower bound of $\mathsf{CP}_{N}\left(\lambda\right)$
as
\begin{align}
\mathsf{CP}_{N}\left(\lambda\right)> & \mathsf{CP}_{N}^{\mathrm{L}}\left(\lambda\right)\nonumber \\
= & \mathbb{E}_{r_{0}\in\left[R_{N-1},\infty\right)}\left[e^{-2\pi\lambda\tau\omega_{1}\left(\tau,\alpha_{N-1}\right)\left(r_{0}^{2}+\Delta h^{2}\right)}\right]\nonumber \\
\overset{\left(\mathrm{a}\right)}{=} & \frac{e^{-\pi\lambda\left[R_{N-1}^{2}+2\tau\omega_{1}\left(\tau,\alpha_{N-1}\right)\left(R_{N-1}^{2}+\Delta h^{2}\right)\right]}}{1+2\tau\omega_{1}\left(\tau,\alpha_{N-1}\right)},\label{eq:bound proof 3-1}
\end{align}
where (a) is due to the PDF of $r_{0}$ given in (\ref{eq:PDF of r0}).
Therefore, it can be shown that $\exists\frac{1}{1+2\tau\omega_{1}\left(\tau,\alpha_{N-1}\right)}>0$,
$\forall\lambda>0$, 
\begin{align}
\left|\mathsf{CP}_{N}^{\mathrm{L}}\left(\lambda\right)\right| & \geq\frac{e^{-\pi\lambda\left[R_{N-1}^{2}+2\tau\omega_{1}\left(\tau,\alpha_{N-1}\right)\left(R_{N-1}^{2}+\Delta h^{2}\right)\right]}}{1+2\tau\omega_{1}\left(\tau,\alpha_{N-1}\right)}.\label{eq:bound proof 4}
\end{align}
Hence, $\mathsf{CP}_{N}^{\mathrm{L}}\left(\lambda\right)=\Omega\left(e^{-\pi\lambda\left[R_{N-1}^{2}+2\tau\omega_{1}\left(\tau,\alpha_{N-1}\right)\left(R_{N-1}^{2}+\Delta h^{2}\right)\right]}\right)$
holds true. 

In the following, we analyze the upper bound of $\mathsf{CP}_{N}\left(\lambda\right)$.
When $r_{0}\in\left[R_{n},R_{n+1}\right)$ or equivalently $d_{0}\in\left[\bar{R}_{n},\bar{R}_{n+1}\right)$
$\left(n=0,1,\ldots,N-2\right)$, $s_{N}=\frac{\tau d_{0}^{\alpha_{n}}}{PK_{n}}$
holds. As such, $\int_{d_{0}}^{\infty}x\left(1-\frac{1}{1+s_{N}Pl_{N}\left(x\right)}\right)dx$
in the first term of (\ref{eq:bound proof 1}) can be manipulated
as
\begin{align}
 & \int_{d_{0}}^{\infty}x\left(1-\frac{1}{1+s_{N}Pl_{N}\left(x\right)}\right)dx\nonumber \\
\overset{\left(\mathrm{a}\right)}{>} & \int_{\bar{R}_{N-1}}^{\infty}x\left(1-\frac{1}{1+\frac{\tau K_{N-1}}{K_{n}d_{0}^{-\alpha_{n}}}x^{-\alpha_{N-1}}}\right)dx\nonumber \\
= & \frac{\tau K_{N-1}\bar{R}_{N-1}^{2-\alpha_{N-1}}d_{0}^{\alpha_{n}}}{K_{n}\left(\alpha_{N-1}-2\right)}\omega_{1}\left(\frac{\tau K_{N-1}d_{0}^{\alpha_{n}}}{K_{n}R_{N-1}^{\alpha_{N-1}}},\alpha_{N-1}\right)\nonumber \\
\overset{\left(\mathrm{b}\right)}{>} & \frac{\tau K_{N-1}\bar{R}_{N-1}^{2-\alpha_{N-1}}\Delta h^{\alpha_{n}}}{K_{n}\left(\alpha_{N-1}-2\right)}\omega_{1}\left(\frac{\tau K_{N-1}}{K_{n}},\alpha_{N-1}\right)\nonumber \\
= & q_{1}\left(n\right),\label{eq:bound proof 4 - 1}
\end{align}
where (a) follows due to $d_{0}<\bar{R}_{N-1}$, and (b) follows because
$d_{0}>\Delta h$, $d_{0}^{\alpha_{n}}<R_{N-1}^{\alpha_{N-1}}$ and
$\omega_{1}\left(x,\alpha_{N-1}\right)$ is a decreasing function
of $x$ (see Lemma \ref{lemma: hypergeometric function}). Using (\ref{eq:bound proof 4 - 1})
and the PDF of $r_{0}$ in (\ref{eq:PDF of r0}), we have 
\begin{align}
 & \mathbb{E}_{r_{0}\in\left[R_{0},R_{N-1}\right)}\left[e^{-2\pi\lambda\int_{d_{0}}^{\infty}x\left(1-\frac{1}{1+s_{N}Pl_{N}\left(x\right)}\right)dx}\right]\nonumber \\
< & \stackrel[n=0]{N-2}{\sum}\mathbb{E}_{r_{0}\in\left[R_{n},R_{n+1}\right)}\left[e^{-2\pi\lambda q_{1}\left(n\right)}\right]\nonumber \\
= & \stackrel[n=0]{N-2}{\sum}e^{-2\pi\lambda q_{1}\left(n\right)}\left(e^{-\pi\lambda R_{n}^{2}}-e^{-\pi\lambda R_{n+1}^{2}}\right).\label{eq:bound proof 5}
\end{align}

When $r_{0}\in\left[R_{N-1},\infty\right)$, the second term of (\ref{eq:bound proof 1})
is already given by $\mathsf{CP}_{N}^{\mathrm{L}}\left(\lambda\right)$
in (\ref{eq:bound proof 3-1}). Hence, it is easy to obtain
\begin{align}
\mathsf{CP}_{N}\left(\lambda\right)< & \stackrel[n=0]{N-2}{\sum}e^{-2\pi\lambda q_{1}\left(n\right)}\left(e^{-\pi\lambda R_{n}^{2}}-e^{-\pi\lambda R_{n+1}^{2}}\right)+\mathsf{CP}_{N}^{\mathrm{L}}\left(\lambda\right)\nonumber \\
< & \stackrel[n=0]{N-2}{\sum}e^{-2\pi\lambda q_{1}\left(n\right)}e^{-\pi\lambda R_{n}^{2}}+\mathsf{CP}_{N}^{\mathrm{L}}\left(\lambda\right)\nonumber \\
< & \stackrel[n=0]{N-2}{\sum}e^{-2\pi\lambda q_{1}\left(n\right)}+e^{-\pi\lambda R_{N-1}^{2}}\nonumber \\
= & \mathsf{CP}_{N}^{\mathrm{U}}\left(\lambda\right).\label{eq:bound proof 6}
\end{align}
In (\ref{eq:bound proof 6}), If $n\in\mathbb{C}$ $\left(\mathbb{C}=\left\{ 0,1,\ldots,N-2\right\} \right)$,
which enables $2q_{1}\left(n\right)>R_{N-1}^{2}$, then the inequality
$e^{-2\pi\lambda q_{1}\left(n\right)}<e^{-\pi\lambda R_{N-1}^{2}}$
holds. Then, $\mathsf{CP}_{N}^{\mathrm{U}}\left(\lambda\right)$ in
(\ref{eq:bound proof 6}) turns into 
\begin{align*}
\mathsf{CP}_{N}^{\mathrm{U}}\left(\lambda\right)=\stackrel[n=0]{N-2}{\sum}e^{-2\pi\lambda q_{1}\left(n\right)}+e^{-\pi\lambda R_{N-1}^{2}}< & Ne^{-\pi\lambda R_{N-1}^{2}},
\end{align*}
which indicates that $\exists N>0$, $\forall\lambda>0$,
\begin{align}
\left|\mathsf{CP}_{N}^{\mathrm{U}}\left(\lambda\right)\right|< & Ne^{-\pi\lambda R_{N-1}^{2}}.\label{eq:scaling proof 2}
\end{align}

If $n\in\mathbb{C}^{\dagger}$ $\left(\mathbb{C}\subseteq\left\{ 0,1,\ldots,N-2\right\} \right)$,
which enables $2q_{1}\left(n\right)\leq R_{N-1}^{2}$, then we denote
$n=N^{\dagger}$, which makes $e^{-2\pi\lambda q_{1}\left(N^{\dagger}\right)}\geq e^{-2\pi\lambda q_{1}\left(n\right)}$
$\left(0\leq n\leq N-2\right)$. It is apparent that $e^{-2\pi\lambda q_{1}\left(N^{\dagger}\right)}\geq e^{-\pi\lambda R_{N-1}^{2}}$
holds as well. Then, we have
\begin{align*}
\stackrel[n=0]{N-2}{\sum}e^{-2\pi\lambda q_{1}\left(n\right)}+e^{-\pi\lambda R_{N-1}^{2}}< & Ne^{-2\pi\lambda q_{1}\left(N^{\dagger}\right)}.
\end{align*}
In this case, $\exists N>0$, $\forall\lambda>0$,
\begin{align}
\left|\mathsf{CP}_{N}^{\mathrm{U}}\left(\lambda\right)\right| & <Ne^{-2\pi\lambda q_{1}\left(N^{\dagger}\right)}.\label{eq:scaling proof 3}
\end{align}
Combining (\ref{eq:scaling proof 2}) and (\ref{eq:scaling proof 3}),
$\mathsf{CP}_{N}^{\mathrm{U}}\left(\lambda\right)=\mathcal{O}\left(e^{-\pi\lambda R_{N-1}^{2}}\right)$
or $\mathsf{CP}_{N}^{\mathrm{U}}\left(\lambda\right)=\mathcal{O}\left(e^{-2\pi\lambda q_{1}\left(N^{\dagger}\right)}\right)$
holds true. 

According to the above proof for the scaling laws of $\mathsf{CP}_{N}^{\mathrm{U}}\left(\lambda\right)$
and $\mathsf{CP}_{N}^{\mathrm{L}}\left(\lambda\right)$, it is easy
to show that there exists a constant $\kappa$, which makes $\mathsf{CP}_{N}\left(\lambda\right)$
scale with $\lambda$ as $e^{-\kappa\lambda}$. Therefore, based on
the definition of ST in (\ref{eq: define ST}), $\mathsf{ST}_{N}\left(\lambda\right)$
scales with $\lambda$ as $\lambda e^{-\kappa\lambda}$.

\subsection{Proof for Theorem \ref{theorem: CP and ST scaling law}\label{subsec:Proof for Theorem condition}}

Substituting the special case of CP $\left(N=1\right)$ in (\ref{eq:CP general})
into (\ref{eq: define CP with QoS Req}), we have $\frac{1}{1+C_{1}}\exp\left(-\pi\lambda C_{1}\bigtriangleup h^{2}\right)>\varepsilon$.
Through easy manipulation, the following inequality can be obtained
\begin{align}
\lambda<-\frac{\ln\left[\varepsilon\left(1+C_{1}\right)\right]}{\pi C_{1}\bigtriangleup h^{2}} & .\label{eq:proof for theorem condition}
\end{align}

To make the inequality in (\ref{eq:proof for theorem condition})
valid, we have to guarantee $\ln\left[\varepsilon\left(1+C_{1}\right)\right]<0$.
Hence, the proof is complete.

\bibliographystyle{IEEEtran}
\bibliography{ref_BPM}

\begin{thebibliography}{10}
\providecommand{\url}[1]{#1}
\csname url@samestyle\endcsname
\providecommand{\newblock}{\relax}
\providecommand{\bibinfo}[2]{#2}
\providecommand{\BIBentrySTDinterwordspacing}{\spaceskip=0pt\relax}
\providecommand{\BIBentryALTinterwordstretchfactor}{4}
\providecommand{\BIBentryALTinterwordspacing}{\spaceskip=\fontdimen2\font plus
\BIBentryALTinterwordstretchfactor\fontdimen3\font minus
  \fontdimen4\font\relax}
\providecommand{\BIBforeignlanguage}[2]{{%
\expandafter\ifx\csname l@#1\endcsname\relax
\typeout{** WARNING: IEEEtran.bst: No hyphenation pattern has been}%
\typeout{** loaded for the language `#1'. Using the pattern for}%
\typeout{** the default language instead.}%
\else
\language=\csname l@#1\endcsname
\fi
#2}}
\providecommand{\BIBdecl}{\relax}
\BIBdecl

\bibitem{Network_densification_Ref1}
N.~Bhushan, J.~Li, D.~Malladi, R.~Gilmore, D.~Brenner, A.~Damnjanovic, R.~T.
  Sukhavasi, C.~Patel, and S.~Geirhofer, ``Network densification: the dominant
  theme for wireless evolution into {5G},'' \emph{IEEE Commun. Mag.}, vol.~52,
  no.~2, pp. 82--89, Feb. 2014.

\bibitem{Ref_F_Yu_small_cell}
N.~Zhao, X.~Liu, F.~R. Yu, M.~Li, and V.~C.~M. Leung, ``Communications,
  caching, and computing oriented small cell networks with interference
  alignment,'' \emph{IEEE Commun. Mag.}, vol.~54, no.~9, pp. 29--35, Sep. 2016.

\bibitem{Ref_F_Yu_SC}
L.~Chen, F.~R. Yu, H.~Ji, B.~Rong, X.~Li, and V.~C.~M. Leung, ``Green
  full-duplex self-backhaul and energy harvesting small cell networks with
  massive {MIMO},'' \emph{IEEE J. Sel. Areas Commun.}, vol.~34, no.~12, pp.
  3709--3724, Dec 2016.

\bibitem{UDN_benefit_Qualcomm}
\BIBentryALTinterwordspacing
{Qualcomm Technologies, Inc.}, ``Enabling hyper-dense small cell deployments
  with {UltraSON},'' Tech. Rep., Feb. 2014. [Online]. Available:
  \url{https://www.qualcomm.com/media/documents/files/enabling-hyper-dense-small-cell-deployments-with-ultrason.pdf}
\BIBentrySTDinterwordspacing

\bibitem{UPM_Ref_Original}
J.~G. Andrews, F.~Baccelli, and R.~K. Ganti, ``A tractable approach to coverage
  and rate in cellular networks,'' \emph{IEEE Trans. Commun.}, vol.~59, no.~11,
  pp. 3122--3134, Nov. 2011.

\bibitem{UPM_Ref_K_tier}
H.~S. Dhillon, R.~K. Ganti, F.~Baccelli, and J.~G. Andrews, ``Modeling and
  analysis of {K-Tier} downlink heterogeneous cellular networks,'' \emph{IEEE
  J. Sel. Areas Commun.}, vol.~30, no.~3, pp. 550--560, Apr. 2012.

\bibitem{MUPM_Ref_LOS_NLOS}
C.~Galiotto, N.~K. Pratas, N.~Marchetti, and L.~Doyle, ``A stochastic geometry
  framework for {LOS/NLOS} propagation in dense small cell networks,'' in
  \emph{Proc. IEEE ICC}, London, UK, June. 2015, pp. 2851--2856.

\bibitem{MUPM_Ref_Original}
X.~Zhang and J.~G. Andrews, ``Downlink cellular network analysis with
  multi-slope path loss models,'' \emph{IEEE Trans. Commun.}, vol.~63, no.~5,
  pp. 1881--1894, May. 2015.

\bibitem{MUPM_Ref_LOS_Conf}
M.~Ding, D.~L\'{o}pez-P\'{e}rez, G.~Mao, P.~Wang, and Z.~Lin, ``Will the area
  spectral efficiency monotonically grow as small cells go dense?'' in
  \emph{Proc. IEEE GLOBECOM}, SANDIEGO, CA, Dec. 2015, pp. 1--7.

\bibitem{MUPM_Ref_LOS_Journal}
M.~Ding, P.~Wang, D.~L\'{o}pez-P\'{e}rez, G.~Mao, and Z.~Lin, ``Performance
  impact of {LoS} and {NLoS} transmissions in dense cellular networks,''
  \emph{IEEE Trans. Wireless Commun.}, vol.~15, no.~3, pp. 2365--2380, Mar.
  2016.

\bibitem{MUPM_Ref_K_dimension}
A.~K. Gupta, X.~Zhang, and J.~G. Andrews, ``{SINR} and throughput scaling in
  ultradense urban cellular networks,'' \emph{IEEE Wireless Commun. Lett.},
  vol.~4, no.~6, pp. 605--608, Dec. 2015.

\bibitem{book_stochastic_geometry}
D.~Stoyan, W.~S. Kendall, J.~Mecke, and L.~Ruschendorf, \emph{{Stochastic
  geometry and its applications}}.\hskip 1em plus 0.5em minus 0.4em\relax Wiley
  Chichester, 1995, vol.~2.

\bibitem{Ref_SBPM}
J.~Liu, M.~Sheng, L.~Liu, and J.~Li, ``Effect of densification on cellular
  network performance with bounded pathloss model,'' \emph{IEEE Commun. Lett.},
  vol.~21, no.~2, pp. 346--349, Feb. 2017.

\end{thebibliography}

\end{document}